\documentclass[aps,prl,onecolumn,nofootinbib,secnumbers,secnumarabic,longbibliography]{revtex4-2}

\usepackage{amsmath,amssymb,amsthm,mathtools}
\usepackage[most]{tcolorbox}
\usepackage{xcolor}
\usepackage{enumitem}
\usepackage{microtype}

\usepackage[hidelinks,bookmarksopen=true,bookmarksnumbered=true]{hyperref}

\usepackage[nameinlink,capitalise]{cleveref}

\newcommand{\Vol}{\mathcal{V}}

\makeatletter
\@ifundefined{Ce}{}{}
\@ifundefined{Ctilde}{}{}
\makeatother

\theoremstyle{plain}

\theoremstyle{definition}

\theoremstyle{remark}

\tcbset{
  enhanced jigsaw,
  breakable,
  boxrule=0.6pt,
  colframe=black!40,
  colback=black!3,
  arc=1mm,
  left=0.9em,right=0.9em,top=0.7em,bottom=0.7em
}

\makeatletter

\def\p@section{}
\def\p@subsection{}
\makeatother

\begin{document}

\title{Geometric Characterization of Anisotropic Correlations \\ via Mutual Information Tomography}
\author{Beau Leighton-Trudel}
\affiliation{Independent Researcher, Saint Petersburg, Florida, USA}
\date{\today}

\begin{abstract}
Characterizing anisotropic correlations in quantum and statistical systems requires a coordinate-invariant framework. We introduce a geometric map based on the local informational line element, calibrated by the Euclidean benchmark scale $C_{\text{vac}}$:
\[
ds^2 = \frac{C_{\text{vac}}}{I(x, x+\epsilon)}.
\]
We prove that this map yields a smooth Riemannian structure ($g_{ij}$) if and only if the short-distance mutual information (MI) follows the anisotropic inverse-quadratic law (local exponent $X_{\text{loc}}=2$). A key insight is that anisotropy is necessary to activate tensor geometry; isotropic MI forces conformal flatness ($g_{ij} \propto \delta_{ij}$), suppressing shear degrees of freedom. We employ a parameterization-invariant unimodular split, $g_{ij} = \Vol^{2/D}\gamma_{ij}$, which rigorously separates local density fluctuations (volume $\Vol$) from directional anisotropy (shape/shear $\gamma_{ij}$). We introduce ``MI Tomography,'' an operational protocol to reconstruct these geometric components from finite directional measurements. The protocol is validated using the equal-time ground state of an anisotropic 2D quantum harmonic lattice (massless relativistic scalar) on a torus, where the reconstructed shape tensor $\gamma_{ij}$ quantitatively recovers the physical coupling anisotropy. We work strictly in the local, fixed-coarse-graining $X_{\text{loc}}=2$ branch; the line element is used solely to extract the local kinematic structure (the local metric tensor), deferring global distance claims.
\end{abstract}

\maketitle

\section{INTRODUCTION}

Anisotropy is a ubiquitous feature of quantum and statistical systems, arising from directional couplings, competing interactions, or the emergence of ordered phases such as nematic states~\cite{Oganesyan2001Nematic,Fradkin2010NematicReview}. Characterizing these directional correlations in a rigorous, coordinate-invariant manner is essential for understanding the underlying physics. Geometric tools offer a powerful framework for analyzing the structure of correlations, translating informational measures into spatial intuition~\cite{CaoCarrollMichalakis2017}.

Recent work introduced a calibrated informational distance, $d_E = K_0/\sqrt{I}$, to probe the global metric structure of correlations in one-dimensional quantum chains, with the normalization $C_{\mathrm{vac}} = K_0^2$ uniquely fixed by the Euclidean benchmark: isotropic scaling $I(r)\propto r^{-2}$ (exponent $X=2$) must map to $d_E(r)\propto r$~\cite{LeightonTrudel2025Metricity}. This framework established that for power-law correlations ($I \propto r^{-X}$), a coherent metric structure emerges if and only if $0 < X \le 2$~\cite{LeightonTrudel2025Metricity, LeightonTrudel2025GSR}. Furthermore, a smooth (Riemannian) geometry emerges if and only if the scaling saturates the benchmark, $X=2$ (SM~\S S2). However, this isotropic realization presents a crucial limitation: it necessarily yields a strictly conformally flat metric, $g_{ij}(x) = \Phi(x)\delta_{ij}$. Such a geometry captures only local density variations (via $\Phi(x)$) and suppresses the tensor degrees of freedom (shape and shear) that characterize anisotropic systems and many ordered phases.

In this Letter, we demonstrate that anisotropy in the mutual information is the essential ingredient required to activate the full tensor geometry. We establish a parameterization-invariant map from the structure of short-distance MI to a local spatial metric $g_{ij}(x)$. We prove a central kinematic equivalence: applying the postulate as a local line element, $ds^2 = C_{\text{vac}}/I(x, x+\epsilon)$, yields a smooth Riemannian manifold if and only if the local MI ($X_{\text{loc}}=2$) follows the anisotropic inverse-quadratic law:
\begin{equation}
I(x, x+\epsilon) \approx \frac{\tilde{C}(x)}{h_{ij}(x)\epsilon^i\epsilon^j},
\label{eq:anisotropic_law}
\end{equation}
where $h_{ij}(x)$ is a positive-definite tensor encoding directional dependence and $\tilde{C}(x)$ is the local MI magnitude. This uniquely induces the metric:
\begin{equation}
g_{ij}(x) = \Phi(x)h_{ij}(x), \quad \Phi(x) = \frac{C_{\text{vac}}}{\tilde{C}(x)}.
\label{eq:metric_map}
\end{equation}

To isolate the physical degrees of freedom from the redundancy inherent in the factorization of Eq.~(\ref{eq:anisotropic_law}), we employ a parameterization-invariant unimodular split, $g_{ij} = \Vol^{2/D}\gamma_{ij}$. This rigorously separates the geometry into a volume density ($\Vol$) and a unimodular shape/shear tensor ($\gamma_{ij}$). We provide an operational protocol, ``MI Tomography,'' to reconstruct these components from finite directional MI measurements and validate it using an anisotropic 2D quantum harmonic lattice (massless relativistic free scalar) model.

\textit{Scope and Operational Regime.}---To ensure rigor, we emphasize the strict scope of this analysis. We operate within a fixed coarse-graining regime (cell size $l$) and restrict our analysis to the short-distance branch where the local scaling exponent is $X_{\text{loc}}=2$. This universality class includes, for example, critical 1D chains described by conformal field theory~\cite{CalabreseCardy2004,CalabreseCardy2009} and 2D massless relativistic scalar fields, but excludes systems such as free 2D fermions and 3D scalars where $X>2$. In higher-dimensional interacting critical points with anomalous dimension $\eta>0$ (for example, the $(2+1)$-dimensional Ising universality class), standard scaling gives $\langle\phi(0)\phi(r)\rangle \sim r^{-(D-1+\eta)}$%
~\cite{Cardy1996Scaling,Sachdev2011QPT}
and, in the weak-correlation regime, $I(r)\propto r^{-X}$ with
$X = 2(D-1+\eta)>2$~\cite{Wolf2008AreaLaw,PeschelEisler2009}. In such cases the postulate $d_E^2 = C_{\text{vac}}/I$ lies outside the Riemannian branch: globally $d_E(r)\propto r^{X/2}$ violates the triangle inequality for $X>2$, and locally there is no quadratic (Riemannian) line element~\cite{LeightonTrudel2025Metricity,LeightonTrudel2025GSR}. Interacting fixed points with $\eta>0$ are therefore explicitly out of scope in the present work. This is strictly a \textit{local kinematic analysis} of static correlations. We employ the line element $ds^2 = C_{\text{vac}}/I$ solely to extract the local metric tensor $g_{ij}(x)$. We make no claims that the integrated geodesic distances of $g$ equal the global pairwise distance $d_E = \sqrt{C_{\text{vac}}/I}$ at finite separations, except in the homogeneous, isotropic $X=2$ benchmark.

\section{KINEMATICS OF ANISOTROPIC INFORMATION}

We now analyze the local structure of correlations by applying the informational distance postulate at infinitesimal separations. This establishes the bridge between the short-distance structure of mutual information and the local differential geometry.

\textit{The Local Map and Riemannian Condition.}---At fixed coarse-graining $l$, we define the local line element:
\begin{equation}
ds^2(x;\epsilon) := \frac{C_{\text{vac}}}{I(x, x+\epsilon)}.
\label{eq:line_element}
\end{equation}
As established in the isotropic case (SM~\S S2), this definition yields a smooth Riemannian structure if and only if the local MI scaling exponent saturates the Euclidean benchmark, $X_{\text{loc}}=2$. We operate strictly within this regime.

\textit{The Anisotropic Inverse-Quadratic Law.}---The most general realization of the $X_{\text{loc}}=2$ condition is the anisotropic inverse-quadratic profile:
\begin{equation}
I(x, x+\epsilon) = \frac{\tilde{C}(x)}{h_{ij}(x)\epsilon^i\epsilon^j}[1+R(x,\epsilon)],
\label{eq:anisotropic_law_main}
\end{equation}
where $\tilde{C}(x)>0$ is the local magnitude, $h_{ij}(x)$ is a positive-definite, symmetric tensor encoding the directional dependence, and the remainder scales as $R(x,\epsilon)=o(1)$. Applying the line element (\ref{eq:line_element}) to this profile yields the emergent metric $g_{ij}(x) = \Phi(x)h_{ij}(x)$, with $\Phi(x) = C_{\text{vac}}/\tilde{C}(x)$.

\textit{The Kinematic Equivalence.}---This specific profile is not merely sufficient; it is necessary. We now state the central theoretical result establishing the equivalence between this MI structure and the emergent geometry.

\textbf{Theorem 1 (Kinematic Equivalence).} \textit{Let $d^2(x,x+\epsilon) := C_{\text{vac}}/I(x,x+\epsilon)$. A smooth, positive-definite Riemannian metric $g_{ij}(x)$ emerges locally if and only if the short-distance MI follows the anisotropic inverse-quadratic law (Eq.~(\ref{eq:anisotropic_law_main})). In this case, the induced metric is uniquely determined by $g_{ij}(x) = \Phi(x)h_{ij}(x)$.}

\textit{Proof Sketch.} (i) Sufficiency: Substituting Eq.~(\ref{eq:anisotropic_law_main}) into Eq.~(\ref{eq:line_element}) directly yields $ds^2 = g_{ij}(x)\epsilon^i\epsilon^j + o(|\epsilon|^2)$, identifying the metric. (ii) Necessity: If $d^2$ admits a positive-definite quadratic 2-jet, $d^2 = g_{ij}\epsilon^i\epsilon^j + o(|\epsilon|^2)$, then $I = C_{\text{vac}}/d^2$ necessarily takes the inverse-quadratic form. (Full details in SM~\S S3.5).

\textit{Parameterization Redundancy and the Unimodular Split.}---While the map $g=\Phi h$ correctly identifies the metric, the factorization of $I$ into inputs $(\tilde{C}, h)$ admits a simple redundancy. The observable MI (and thus $ds^2$) is invariant under the joint rescaling:
\begin{equation}
\mathcal{R}_\sigma: (h, \tilde{C}) \mapsto (\sigma^2 h, \sigma^2 \tilde{C}), \quad \sigma>0.
\end{equation}
This redundancy concerns only the factorization of $I$ and is distinct from coordinate transformations (diffeomorphisms). To isolate the parameterization-invariant physical degrees of freedom—volume versus shape/shear—we employ the unimodular split.

From any representative $h$, we define the unimodular shape density $\gamma_{ij}$ and the volume density $\Vol$:
\begin{align}
\gamma_{ij} &:= (\det h)^{-1/D}h_{ij} \quad (\det \gamma \equiv 1), \\
\Vol &:= \left(\frac{C_{\text{vac}}}{\tilde{C}}\right)^{D/2}\sqrt{\det h}.
\end{align}
The metric is then expressed invariantly as
\begin{equation}
g_{ij} = \Vol^{2/D}\gamma_{ij}.
\label{eq:unimodular_split}
\end{equation}
The field $\gamma$ carries the pure anisotropy (shape/shear), while $\Vol$ is precisely the volume element, $\Vol = \sqrt{\det g}$. We will refer to the Riemannian structure $(\Vol(x), \gamma_{ij}(x))$ induced by short-distance mutual information as the local correlation geometry of the state.

\textit{Consistency.}---This local map is self-consistent. By the definition of the exponential map and the Gauss Lemma~\cite{Lee2018Riemannian}, the squared geodesic distance along a radial path is exactly $d_{\text{geo}}^2(x, \exp_x(\epsilon)) = g_{ij}(x)\epsilon^i\epsilon^j$. This confirms that the metric $g$ is entirely determined by the local quadratic relationship established by the MI profile (SM~\S S3.3). Curvature is encoded in the spatial derivatives of the fundamental fields $\Vol$ and $\gamma$.

\section{OPERATIONAL PROTOCOL: MI TOMOGRAPHY}

The kinematic map established in Section II provides the theoretical foundation; we now present an operational protocol, MI Tomography, to reconstruct the local metric $g_{ij}(x)$ and its invariant components $(\Vol, \gamma)$ from a finite set of directional MI measurements.

\textit{The Observable.}---The reconstruction relies on the observable derived from the inverse mutual information:
\begin{equation}
y(\epsilon) := \frac{1}{I(x, x+\epsilon)}.
\end{equation}
In the $X_{\text{loc}}=2$ regime, the line element postulate ensures that $y(\epsilon)$ is proportional to the squared emergent distance. This yields the crucial quadratic relationship (Eq.~(\ref{eq:anisotropic_law_main})):
\begin{equation}
y(\epsilon) = \frac{1}{C_{\text{vac}}}g_{ij}(x)\epsilon^i\epsilon^j + \text{higher order}.
\label{eq:y_quadratic}
\end{equation}
The components of the metric $g_{ij}(x)$ are thus encoded in the quadratic coefficients of the observable $y(\epsilon)$.

\textit{The Reconstruction Protocol.}---The structure of a quadratic form allows the coefficients to be extracted via the polarization identity~\cite{HornJohnson2013MatrixAnalysis}. By sampling $y(\epsilon)$ at a single scale factor $\rho$ along a sufficient set of directions, we can determine the $D(D+1)/2$ independent components of the metric.

To mitigate numerical noise and cancel odd-order contributions in the remainder of Eq.~(\ref{eq:y_quadratic}), we utilize the symmetrized observable: $\bar{y}(v) := \frac{1}{2}[y(v) + y(-v)]$. A minimal, closed-form estimator using the coordinate axes $\{e_i\}$ and pair-sums $\{e_i+e_j\}$ directions is given by:
\begin{align}
g_{ii}(x) &= \frac{C_{\text{vac}}}{\rho^2}\bar{y}(\rho e_i), \label{eq:tomography_diag} \\
g_{ij}(x) &= \frac{C_{\text{vac}}}{2\rho^2}[\bar{y}(\rho(e_i+e_j)) - \bar{y}(\rho e_i) - \bar{y}(\rho e_j)], \quad (i<j). \label{eq:tomography_offdiag}
\end{align}
This reconstruction is inherently parameterization-invariant, as the input $\bar{y}$ is unchanged under the redundancy $\mathcal{R}_\sigma$.

\begin{figure}[t]
\centering
\fbox{\begin{minipage}{0.95\columnwidth}
\textbf{Recipe: MI Tomography in 5 Steps}
\begin{enumerate}
    \item \textbf{Setup:} Fix a point $x$ and coarse-graining scale $l$.
    \item \textbf{Sampling:} Choose a small radius $\rho$ ($l \ll \rho$) and directions (axes and axis-pairs).
    \item \textbf{Measurement:} Measure $I(x, x\pm\epsilon)$, form $y=1/I$, and compute the symmetrized observable $\bar{y}(\epsilon)$.
    \item \textbf{Reconstruction:} Compute $\hat{g}_{ij}$ via Eqs.~(\ref{eq:tomography_diag})--(\ref{eq:tomography_offdiag}) (or use regression for oversampled data; SM~\S S5.2).
    \item \textbf{Output:} Regularize $\hat{g}$ to a nearby positive-definite matrix and extract the invariant components $\hat{\Vol}, \hat{\gamma}$.
\end{enumerate}
\end{minipage}}
\end{figure}

\textit{Diagnostics and Validation.}---The validity of this reconstruction rests fundamentally on the assumption that $y(\epsilon)$ is locally quadratic (the $X_{\text{loc}}=2$ condition). This assumption must be verified operationally using two differential tests (DT) (SM~\S S5.0):

\begin{itemize}
    \item \textbf{DT1 (Hessian Identity):} The metric must correspond to the Hessian of the observable at the origin: $g_{ij} = \frac{C_{\text{vac}}}{2}\partial_{\epsilon^i}\partial_{\epsilon^j}y(\epsilon)|_{\epsilon=0}$.
    \item \textbf{DT2 (Polarization Test):} The quadratic form must satisfy the polarization identity up to the finite-radius remainder envelope: $y(u+v)+y(u-v)-2y(u)-2y(v) = \text{small}$.
\end{itemize}
Systematic failure of these tests signals a departure from the Riemannian regime ($X_{\text{loc}} \neq 2$ or non-quadratic leading behavior), indicating the kinematic map is inapplicable.

\textit{Bias and Robustness.}---Reconstruction at a finite radius $\rho$ introduces a systematic bias due to the "higher order" terms in Eq.~(\ref{eq:y_quadratic}). If the remainder in the MI profile (Eq.~(\ref{eq:anisotropic_law_main})) scales as $R(x,\epsilon) = O((|\epsilon|/L_*)^{\alpha})$, the bias in the metric estimate $\hat{g}$ scales as $O((\rho/L_*)^{\alpha})$ (proven in SM~\S S6.1). The use of the symmetrized observable $\bar{y}$ is crucial as it cancels all odd powers of $\epsilon$, potentially improving the effective bias exponent.

In practical applications, the resulting estimate $\hat{g}$ must be physically valid and non-singular. We therefore regularize $\hat{g}$ by projecting it onto a nearby positive-definite matrix in the Frobenius norm (implemented via the nearest-PSD algorithm of Ref.~\cite{Higham2002NearestPSD} together with a small eigenvalue floor $\epsilon>0$ on the spectrum). The final result is then reported using the parameterization-invariant unimodular split
$\hat{\Vol}=\sqrt{\det\hat{g}}$ and
$\hat{\gamma}=\hat{\Vol}^{-2/D}\hat{g}$.
While the closed-form estimator is minimal, sampling additional directions improves stability and allows for robust regression techniques (SM~\S S5.2).

\begin{figure}[t]
    \centering
    \includegraphics[width=\columnwidth]{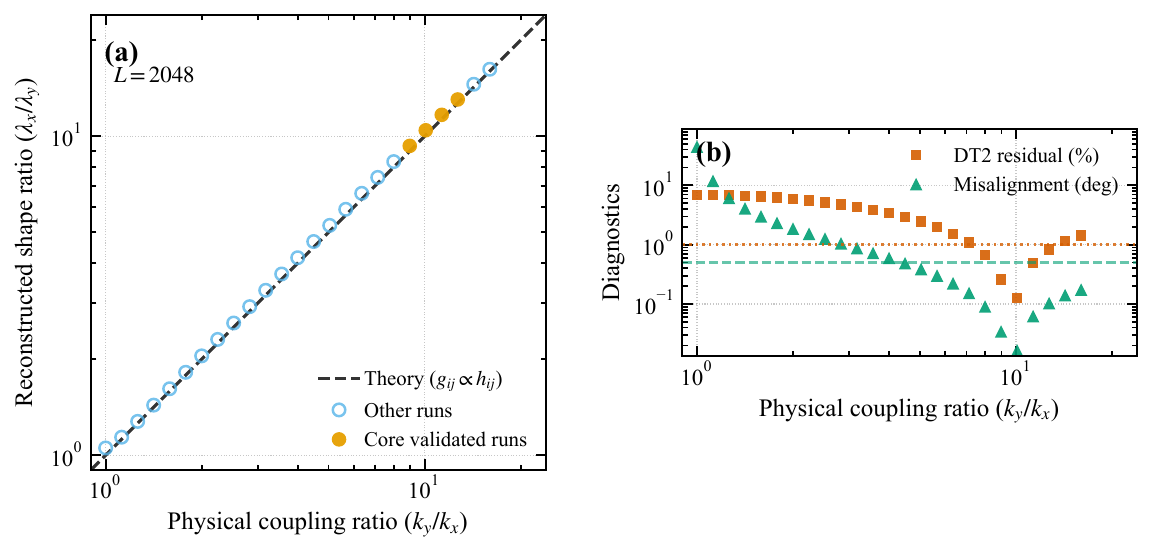}
    \caption{
        Mutual information tomography on the ground state of an anisotropic 2D quantum harmonic lattice (free scalar).
        (a) Reconstructed unimodular shape anisotropy, quantified by the eigenvalue
        ratio $\lambda_x/\lambda_y$ of the shape tensor $\gamma_{ij}$, versus the
        microscopic coupling ratio $k_y/k_x$ for an $L=2048$ lattice with
        $m=10^{-3}$, $\ell=1$, and sampling radius $\rho=50$. Open symbols denote
        all validated runs; filled symbols highlight the ``core'' subset with
        $|X_{\mathrm{loc}}-2|\le 0.05$ and DT2~$\le 1\%$. The dashed line is the
        parameter-free prediction $g_{ij}\propto h_{ij}$ with
        $h_{ij}\propto\mathrm{diag}(1/k_x,1/k_y)$. (b) Diagnostics for the same runs:
        normalized DT2 polarization residual (squares) and eigenframe misalignment
        relative to the lattice axes (triangles), plotted as functions of $k_y/k_x$.
        Horizontal dashed lines indicate the tolerance thresholds DT2~$=1\%$ and
        misalignment~$=0.5^\circ$ used to define the core subset.
    }
    \label{fig:anisotropy_validation}
\end{figure}

\section{APPLICATION: ANISOTROPIC 2D QUANTUM HARMONIC LATTICE}

We validate MI Tomography on a microscopic model where the directional structure of correlations is analytically controlled and the $X_{\text{loc}}=2$ condition is realized.

\textit{Testbed model: anisotropic 2D quantum harmonic lattice on a torus}.---We consider the equal-time ground state of a quadratic (Gaussian) bosonic lattice on a square $L\times L$ torus,
\begin{equation}
H = \frac{1}{2}\sum_{x\in\Lambda}\bigl[p_{x}^{2}+m^{2}q_{x}^{2}\bigr]
+\frac{1}{2}\sum_{x\in\Lambda}\sum_{\mu=x,y}k_{\mu}\bigl(q_{x}-q_{x+\hat{e}_{\mu}}\bigr)^{2},
\label{eq:bosonic_hamiltonian}
\end{equation}
with anisotropy introduced via $k_x\neq k_y$. A small mass $m>0$ acts as an infrared regulator near the massless critical point. In momentum space the low–energy dispersion has the anisotropic relativistic form
$\omega^{2}(\mathbf{q})\simeq m^{2}+k_{x}q_{x}^{2}+k_{y}q_{y}^{2}$, so in the continuum limit this is just the free scalar field in $2+1$ dimensions with anisotropic velocities. We simulate this discretized continuum limit directly using a spectral representation with dispersion \(\omega^{2}(\mathbf{q}) = m^{2} + k_{x} q_{x}^{2} + k_{y} q_{y}^{2}\) evaluated on the discrete momentum grid; the ground state is Gaussian and all mutual informations are computed exactly from the covariance matrix via the symplectic–eigenvalue formula~\cite{Weedbrook2012GaussianQI,PeschelEisler2009}. Details of the spectral solver and MI computation are given in SM~\S S8.

\textit{Terminology note}.---In the probability and statistical mechanics literature the phrase ``2D Gaussian free field'' usually denotes the classical Euclidean massless field with logarithmic covariance, $\langle\phi(x)\phi(0)\rangle\sim -\log r$.~\cite{Sheffield2007GFF}. Here we instead work with the equal-time ground state of a relativistic free scalar (a quantum harmonic lattice) with linear dispersion $\omega\propto|\mathbf{q}|$, which in $D=2$ has $\langle q_{0}q_{r}\rangle\propto 1/r$~\cite{Cardy1996Scaling,Sachdev2011QPT} and hence $I(r)\propto r^{-2}$ (SM~\S S8). To avoid confusion we therefore refer to this testbed simply as an anisotropic quantum harmonic lattice (or free scalar) rather than as the ``2D Gaussian free field.''

\textit{Analytical expectation.}---In a massless relativistic free scalar (Gaussian bosonic field) with $\omega(k)\propto |k|$ in $D$ spatial dimensions the equal–time correlator decays as $\langle q_{0}q_{r}\rangle\propto r^{-(D-1)}$~\cite{Cardy1996Scaling,Sachdev2011QPT}. In $D=2$ this gives $\langle q_{0}q_{r}\rangle\propto r^{-1}$, and in the weak–correlation regime the Gaussian MI satisfies $I\propto\mathrm{Tr}\,C^{2}$~\cite{Wolf2008AreaLaw,PeschelEisler2009}, so $I(r)\propto r^{-2}$ and hence $X_{\text{loc}}=2$. With $k_{x}\neq k_{y}$, the anisotropic dispersion
$\omega^{2}(q)\simeq m^{2}+k_{x}q_{x}^{2}+k_{y}q_{y}^{2}$ deforms this into an inverse–quadratic law with an elliptic radius $r_{h}^{2}=h_{ij}\epsilon^{i}\epsilon^{j}$,
\begin{equation}
I(x,x+\epsilon)\simeq \frac{\tilde C(x)}{h_{ij}(x)\epsilon^{i}\epsilon^{j}},\qquad
h_{ij}\propto \mathrm{diag}\!\left(\frac{1}{k_{x}},\frac{1}{k_{y}}\right),
\label{eq:h_expectation}
\end{equation}
as derived in SM~\S S8.1. In the kinematic map $ds^{2}=C_{\rm vac}/I$ this implies $g_{ij}=\Phi h_{ij}$ with $\Phi(x)=C_{\rm vac}/\tilde C(x)$, so the unimodular shape tensor $\gamma_{ij}$ is proportional to $h_{ij}$ and its eigenvalue ratio is predicted to match the physical anisotropy $k_{y}/k_{x}$.

\textit{Numerical protocol.}---We work at fixed coarse–graining scale $l=1$ (single–site cells) and large system size $L=2048$ with mass $m=10^{-3}$, corresponding to a correlation length $\xi\simeq 1/m\approx 10^{3}$. The MI Tomography protocol (Sec.~III) is implemented at a single sampling radius $\rho=50$, chosen to lie deep in the contact window $l\ll\rho\ll\xi\ll L$; this separation of scales is verified explicitly in the simulations. For each run we measure the mutual information between two congruent single–site cells $A_{\ell}(x)$ and $B_{\ell}(x+\epsilon)$, form $y(\epsilon)=1/I(x,x+\epsilon)$, and compute the symmetrized observable $\bar y(\epsilon)=[y(\epsilon)+y(-\epsilon)]/2$. Using the minimal “axes~+~pair–sum’’ design at $|\epsilon|=\rho$, we reconstruct $\hat g_{ij}$ via Eqs.~(\ref{eq:tomography_diag})–(\ref{eq:tomography_offdiag}), regularize to a positive-definite metric as described in Sec.~III, and extract the invariant unimodular shape $\hat{\gamma}_{ij}$ and volume $\hat{\Vol}=\sqrt{\det \hat{g}}$.%

Before quoting $\hat g_{ij}$, each run must pass three kinematic checks: (i) a scaling analysis of the axis MI, $I(\rho\hat e_{x})\sim \rho^{-X_{\rm loc}}$, which consistently yields $1.99\lesssim X_{\rm loc}\lesssim 2.09$ across the anisotropy sweep, confirming that the data lie in the $X_{\rm loc}\approx 2$ branch; (ii) the DT1 Hessian identity; and (iii) the full DT2 polarization test $y(u+v)+y(u-v)-2y(u)-2y(v)$ evaluated at $u=\rho\hat e_{x}$, $v=\rho\hat e_{y}$ (with $\bar y$), which must remain small compared to the typical magnitude of $y$ (SM~\S S5.0). These diagnostics directly probe the quadraticity of $y(\epsilon)$ and bound the finite–radius bias $O\!\bigl((\rho/L_{\ast})^{\alpha}\bigr)$ from higher–order corrections (SM~\S S6.1).%

\textit{Anisotropy sweep and reconstruction.}---To test the directional response, we vary the coupling anisotropy over two decades while keeping the geometric mean $\sqrt{k_{x}k_{y}}$ fixed. Specifically, we scan $25$ logarithmically spaced ratios $1\leq k_{y}/k_{x}\leq 16$ and, for each, reconstruct the metric at a single point and radius as described above. The resulting unimodular shapes are diagonalized to obtain eigenvalues $(\lambda_{x},\lambda_{y})$ and a principal axis; we define the reconstructed anisotropy ratio $R_{\rm recon}=\lambda_{x}/\lambda_{y}$ and the misalignment angle between the principal eigenvector and the physical $x$–axis. Figure~\ref{fig:anisotropy_validation}(a) compares $R_{\rm recon}$ with the microscopic ratio $k_{y}/k_{x}$ on a log–log scale. Over the entire sweep the points track the identity line with a mean relative error of $\sim\!3.2\%$ and a worst–case deviation of $\simeq 5.4\%$; no fitting parameters are introduced beyond the Euclidean calibration of $C_{\rm vac}$.%

The inset diagnostics in Fig.~\ref{fig:anisotropy_validation}(b) display the DT2 residual (normalized by a typical $\bar y$) and the misalignment angle as functions of $k_{y}/k_{x}$. As expected, the misalignment angle is largest near the isotropic point $k_x = k_y$, where the eigenvalues of the shape tensor become (nearly) degenerate and the principal axis is therefore ill-defined and dominated by numerical noise. Away from this isotropic degeneracy, both quantities rapidly decrease: for $k_y/k_x \gtrsim 9$ the DT2 residual stays at the percent level or below and the eigenframe misalignment is $\lesssim 0.2^\circ$. We highlight this ``core'' subset of runs (filled markers in Fig.~\ref{fig:anisotropy_validation}) where $|X_{\rm loc}-2|\leq 0.05$ and DT2~$\leq 1\%$; within this window the reconstructed anisotropy has a mean relative error $\approx 3.4\%$ with a maximum $\approx 3.9\%$, consistent with the expected finite–radius bias.%

\textit{Significance.}---This microscopic anisotropic 2D quantum harmonic lattice thus realizes the inverse–quadratic short–distance law required by the kinematic framework and provides a stringent, fully controlled test of MI Tomography. Using only local mutual informations between single–site cells at a single radius, the protocol recovers the emergent metric $g_{ij}=\Phi h_{ij}$ and, in particular, the unimodular shape anisotropy set by $k_{y}/k_{x}$, with quantitatively small and well–diagnosed deviations. The agreement between theory and reconstruction in Fig.~\ref{fig:anisotropy_validation}, together with the DT1/DT2 diagnostics and PSD checks, validates MI Tomography as a practical tool for extracting the local correlation geometry of anisotropic statistical systems.

\section{DISCUSSION AND OUTLOOK}

We have established a rigorous kinematic framework for characterizing the geometric structure of anisotropic correlations using mutual information. This local construction forms the kinematic core of a broader correlation geometry framework introduced in Refs.~\cite{LeightonTrudel2025Metricity,LeightonTrudel2025GSR}. By applying the calibrated postulate $ds^2 = C_{\text{vac}}/I$ locally, we derived a necessary and sufficient condition for the emergence of a smooth Riemannian geometry: the anisotropic inverse-quadratic law ($X_{\text{loc}}=2$). This establishes a unique map from the short-distance MI profile to a local metric $g_{ij}(x)$.

A central feature of this framework is the parameterization-invariant unimodular decomposition, $g_{ij} = \Vol^{2/D}\gamma_{ij}$. This provides a rigorous method to separate the physical degrees of freedom within the correlation structure, distinguishing local density variations (volume $\Vol$) from directional anisotropy (shape/shear $\gamma_{ij}$).

This decomposition reveals a key physical insight: isotropy forces a trivial tensor structure, capturing only scalar density variations. Anisotropy in the mutual information is therefore the essential mechanism required to activate the tensor degrees of freedom (shape and shear). This upgrades the geometric characterization from a purely scalar description to a full tensor geometry.

To operationalize this framework, we introduced MI Tomography, a concrete protocol for reconstructing the geometric components from finite directional MI measurements. We established rigorous diagnostics (DT1/DT2) to verify the underlying assumptions and derived bounds on the finite-radius bias. The protocol was validated using an anisotropic 2D quantum harmonic lattice (massless relativistic free scalar), demonstrating its ability to quantitatively recover the physical anisotropy ratio from correlation data alone.

\textit{Outlook.}---This geometric toolkit offers new avenues for analyzing complex many-body systems. The MI Tomography protocol can be directly applied to numerical (e.g., DMRG, tensor network) or experimental data where directional mutual information is accessible. Immediate applications include the characterization of anisotropic critical points in $D>1$, where the shape tensor $\gamma_{ij}$ provides a direct measure of the emergent anisotropy. Furthermore, this framework offers a coordinate-invariant method for identifying and characterizing nematic phases or analyzing the local correlation structure in disordered systems. By providing a rigorous decomposition of volume and shear components, this approach offers a distinct and robust diagnostic for statistical mechanics.

\vspace{1em}
\section*{Code and Data Availability}

All source code, data, and figure-generation scripts used in this work are openly available at:
\href{https://doi.org/10.5281/zenodo.17857589}{doi:\,10.5281/zenodo.17857589}.
This archived release (\texttt{v1.0.1}) corresponds to the results and figures reported in this Letter.

\bibliographystyle{apsrev4-2}
\bibliography{references}

@article{LeightonTrudel2025Metricity,
  author        = {Leighton-Trudel, Beau},
  title         = {Emergent Distance and Metricity of Mutual Information in 1D Quantum Chains},
  journal       = {arXiv},
  year          = {2025},
  eprint        = {2507.09749},
  archivePrefix = {arXiv},
  primaryClass  = {cond-mat.stat-mech},
  doi           = {10.48550/arXiv.2507.09749}
}

@article{LeightonTrudel2025GSR,
  author        = {Leighton-Trudel, Beau},
  title         = {Scaling of a Mutual-Information Distance in One-dimensional Quantum Spin Chains},
  journal       = {arXiv},
  year          = {2025},
  eprint        = {2512.00649},
  archivePrefix = {arXiv},
  primaryClass  = {cond-mat.stat-mech},
  doi           = {10.48550/arXiv.2512.00649}
}

@article{Wolf2008AreaLaw,
  author  = {Wolf, Michael M. and Verstraete, Frank and Hastings, Matthew B. and Cirac, J. Ignacio},
  title   = {Area laws in quantum systems: mutual information and correlations},
  journal = {Phys. Rev. Lett.},
  year    = {2008},
  volume  = {100},
  pages   = {070502},
  doi     = {10.1103/PhysRevLett.100.070502}
}

@article{PeschelEisler2009,
  author  = {Peschel, Ingo and Eisler, Viktor},
  title   = {Reduced density matrices and entanglement entropy in free lattice models},
  journal = {J. Phys. A: Math. Theor.},
  year    = {2009},
  volume  = {42},
  pages   = {504003},
  doi     = {10.1088/1751-8113/42/50/504003}
}

@article{Weedbrook2012GaussianQI,
  author  = {Weedbrook, Christian and Pirandola, Stefano and Garc{\'i}a-Patr{\'o}n, Ra{\'u}l and Cerf, Nicolas J. and Ralph, Timothy C. and Shapiro, Jeffrey H. and Lloyd, Seth},
  title   = {Gaussian quantum information},
  journal = {Rev. Mod. Phys.},
  year    = {2012},
  volume  = {84},
  pages   = {621--669},
  doi     = {10.1103/RevModPhys.84.621}
}

@book{Lee2018Riemannian,
  author    = {Lee, John M.},
  title     = {Introduction to Riemannian Manifolds},
  series    = {Graduate Texts in Mathematics},
  volume    = {176},
  publisher = {Springer},
  address   = {Cham},
  year      = {2018},
  edition   = {2},
  doi       = {10.1007/978-3-319-91755-9}
}

@book{HornJohnson2013MatrixAnalysis,
  author    = {Horn, Roger A. and Johnson, Charles R.},
  title     = {Matrix Analysis},
  publisher = {Cambridge University Press},
  address   = {Cambridge},
  year      = {2013},
  edition   = {2},
  doi       = {10.1017/CBO9781139020414}
}

@article{Higham2002NearestPSD,
  author  = {Higham, Nicholas J.},
  title   = {Computing the nearest correlation matrix -- a problem from finance},
  journal = {IMA J. Numer. Anal.},
  year    = {2002},
  volume  = {22},
  number  = {3},
  pages   = {329--343},
  doi     = {10.1093/imanum/22.3.329}
}

@book{Cardy1996Scaling,
  author    = {Cardy, John},
  title     = {Scaling and Renormalization in Statistical Physics},
  series    = {Cambridge Lecture Notes in Physics},
  volume    = {5},
  publisher = {Cambridge University Press},
  address   = {Cambridge},
  year      = {1996},
  isbn      = {9780521499590}
}

@book{Sachdev2011QPT,
  author    = {Sachdev, Subir},
  title     = {Quantum Phase Transitions},
  publisher = {Cambridge University Press},
  address   = {Cambridge},
  year      = {2011},
  edition   = {2},
  isbn      = {9780521514682}
}

@article{Sheffield2007GFF,
  author  = {Sheffield, Scott},
  title   = {Gaussian free fields for mathematicians},
  journal = {Probab. Theory Relat. Fields},
  year    = {2007},
  volume  = {139},
  number  = {3--4},
  pages   = {521--541},
  doi     = {10.1007/s00440-006-0050-1}
}

@article{Oganesyan2001Nematic,
  author  = {Oganesyan, Vadim and Kivelson, Steven A. and Fradkin, Eduardo},
  title   = {Quantum theory of a nematic Fermi fluid},
  journal = {Phys. Rev. B},
  year    = {2001},
  volume  = {64},
  pages   = {195109},
  doi     = {10.1103/PhysRevB.64.195109}
}

@article{Fradkin2010NematicReview,
  author  = {Fradkin, Eduardo and Kivelson, Steven A. and Lawler, Michael J. and Eisenstein, James P. and Mackenzie, Andrew P.},
  title   = {Nematic Fermi Fluids in Condensed Matter Physics},
  journal = {Annu. Rev. Condens. Matter Phys.},
  year    = {2010},
  volume  = {1},
  pages   = {153--178},
  doi     = {10.1146/annurev-conmatphys-070909-103925}
}

@article{CaoCarrollMichalakis2017,
  author        = {Cao, ChunJun and Carroll, Sean M. and Michalakis, Spyridon},
  title         = {Space from {H}ilbert space: Recovering geometry from bulk entanglement},
  journal       = {Phys. Rev. D},
  year          = {2017},
  volume        = {95},
  pages         = {024031},
  doi           = {10.1103/PhysRevD.95.024031},
  eprint        = {1606.08444},
  archivePrefix = {arXiv},
  primaryClass  = {hep-th}
}

@article{CalabreseCardy2004,
  author        = {Calabrese, Pasquale and Cardy, John},
  title         = {Entanglement Entropy and Quantum Field Theory},
  journal       = {J. Stat. Mech.},
  year          = {2004},
  pages         = {P06002},
  doi           = {10.1088/1742-5468/2004/06/P06002},
  eprint        = {hep-th/0405152},
  archivePrefix = {arXiv},
  primaryClass  = {hep-th}
}

@article{CalabreseCardy2009,
  author        = {Calabrese, Pasquale and Cardy, John},
  title         = {Entanglement entropy and conformal field theory},
  journal       = {J. Phys. A: Math. Theor.},
  year          = {2009},
  volume        = {42},
  pages         = {504005},
  doi           = {10.1088/1751-8113/42/50/504005},
  eprint        = {0905.4013},
  archivePrefix = {arXiv},
  primaryClass  = {cond-mat.stat-mech}
}

\end{document}